# Evolution of Photosynthesis and Biospheric Oxygenation Contingent Upon Nitrogen Fixation?

John W. Grula

*Observatories of the Carnegie Institution of Washington*
*813 Santa Barbara Street*
*Pasadena, CA 91101*
*USA*

## Abstract

How photosynthesis by Precambrian cyanobacteria oxygenated Earth's biosphere remains incompletely understood. Here it is argued that the oxic transition, which took place between approximately 2.3 Gyr and 0.5 Gyr ago, required a great proliferation of cyanobacteria, and this in turn depended on their ability to fix nitrogen via the nitrogenase enzyme system. However, the ability to fix nitrogen was not a panacea, and the rate of biospheric oxygenation may still have been affected by nitrogen constraints on cyanobacterial expansion. Evidence is presented for why cyanobacteria probably have a greater need for fixed nitrogen than other prokaryotes, underscoring the importance of their ability to fix nitrogen. The connection between nitrogen fixation and the evolution of photosynthesis is demonstrated by the similarities between nitrogenase and enzymes critical for the biosynthesis of (bacterio)chlorophyll. It is hypothesized that biospheric oxygenation would not have occurred if the emergence of cyanobacteria had not been preceded by the evolution of nitrogen fixation, and if these organisms had not also acquired the ability to fix nitrogen at the beginning of or very early in their history. The evolution of nitrogen fixation also appears to have been a precondition for the evolution of (bacterio)chlorophyll-based photosynthesis. Given that some form of chlorophyll is obligatory for true photosynthesis, and its light absorption and chemical properties make it a "universal pigment," it may be predicted that the evolution of nitrogen fixation and photosynthesis are also closely linked on other Earth-like planets.

*Key words:* biospheric oxygenation, cyanobacteria, evolution of life, nitrogen fixation, photosynthesis, planetary atmospheres, planetary evolution.

## 1. Introduction

One of the more remarkable transformations of Earth's chemistry by a biological process is the oxygenation of the biosphere, which took place between 2.3 and 0.5 Gyr ago. This event, which apparently unfolded in two main stages, occurred when oxygenic photosynthesis by marine cyanobacteria oxygenated the ocean and increased the oxygen content of the atmosphere by at least several orders of magnitude.  The creation of an oxygen-rich biosphere greatly increased the prevalence of high-energy aerobic metabolisms.  This in turn is generally thought to have led to the appearance of complex life forms, including Ediacaran fauna and most animal phyla, between 600 and 500 million years ago.  The oxic transition also produced the stratospheric ozone layer, which protects life from the most harmful wavelengths of solar ultraviolet radiation.  Thus, prokaryotic life forms dramatically altered Earth's environment and helped create conditions favorable to the evolution of animals, plants, and other complex forms of eukaryotic life (Berman-Frank, Lundgren and Falkowski, 2003; Raymond *et al.*, 2004; Canfield, 2005; Catling et al., 2005; Kump, 2005).

What were the key preconditions for the oxic transition itself?  Certainly one of them was an abundance of water, the substrate from which oxygen is generated during the photosynthetic light reactions (Catling et al., 2005).  But despite the appearance in the marine ecosystem of cyanobacteria and oxygenic photosynthesis by 2.7 Gyr ago and possibly earlier (Des Marais, 2000; Kasting, 2001; Nisbet and Sleep, 2001; Canfield, 2005), it is not fully understood why 400 million years or more elapsed before atmospheric oxygen levels began to rise about 2.3 Gyr ago (Bekker *et al.*, 2004; Canfield, 2005).  While various geochemical explanations have been advanced to explain this (Catling, Zahnle and McKay, 2001; Kasting, 2001; Nisbet and Sleep, 2001; Holland, 2002), a biological factor that also merits consideration is primary production by cyanobacteria, the ultimate oxygen source (Canfield, 2005).  Abundant oxygen "requires a large biological source" (Catling et al., 2005).  Thus, if cyanobacteria had remained a rare life form, oxygenation of the biosphere would probably not have occurred.

When cyanobacteria first appeared about 2.7 Gyr and perhaps earlier, Earth's biosphere had very little oxygen and the processes of respiration and decay were still mainly anaerobic.  Various geochemical sinks for oxygen (such as crustal weathering and combination with volcanic reduced gases) predominated and outpaced cyanobacterial oxygen output for many millions of years (Canfield, 2005; Catling et al., 2005; Condie, 2005).  As cyanobacteria became more widespread, and some geochemical sinks diminished in size, oxygen output eventually exceeded the various sinks and $O_2$ began to accumulate around 2.3 Gyr ago (Holland, 2002; Canfield, 2005; Condie, 2005).  The cyanobacterial biomass could significantly increase only if crucial nutrients such as nitrogen and phosphorus were available in adequate supply (Canfield, 2005).  Here it is hypothesized that the biological process of nitrogen fixation was a key precondition to the proliferation of cyanobacteria and the subsequent oxygenation of the biosphere.  Furthermore, the evolution of nitrogen fixation may also have been a critical precondition for the evolution of photosynthesis itself.

## 2. Fixed Nitrogen Availability and Cyanobacterial Proliferation During the Archean and Proterozoic

While earth's atmosphere has long had abundant amounts of highly unreactive nitrogen gas ($N_2$), environmental fixed nitrogen which is biologically useable was probably a scarce nutrient during much of the Archaean and early Proterozoic (Falkowski, 1997; Kasting and Siefert, 2001; Navarro-Gonzalez, McKay and Mvondo, 2001; Berman-Frank, Lundgren and Falkowski, 2003). This includes the period around 2.7 Gyr ago when cyanobacteria and oxygenic photosynthesis were becoming well established (Des Marais, 2000; Canfield, 2005; Catling et al., 2005). Although nitrogen fixation by lightning would have produced by this time some accumulation of $NO_3^-$ and $NO_2^-$ in the oceans (Mancinelli and McKay, 1988), the period from 2.7 Gyr to 2.3 Gyr ago may still have been a time of especially severe environmental N-limitations on photosynthesis. This would have been the case because slight oxygenation of the ocean photic zone by cyanobacteria would have facilitated anaerobic denitrification at lower depths, resulting in the conversion of nitrate to nitrogen gases. This in turn would have had the effect of greatly reducing the supply of fixed-N compounds in the photic zone, just at the time the growing biomass of cyanobacteria was increasing the demand (Holland, 2002). Any limitations on oxygenic photosynthesis during this period certainly could have contributed to the lag of 400 million years or more between the emergence of cyanobacteria and the initial oxygenation of the atmosphere.

While certain abiotic sources of fixed nitrogen (such as that produced by lightning or delivered to Earth by comets and meteorites) helped life get started, the yields from these sources were likely to have been low and subject to depletion (Mancinelli and McKay, 1988; Falkowski, 1997). Hence, as demand exceeded supply, there was probably strong selection for the ability of at least some organisms to produce their own fixed nitrogen, via the nitrogenase enzyme system, almost as soon as life appeared 3.5 - 3.9 Gyr ago (Falkowski, 1997; Kasting and Siefert, 2001; Navarro-Gonzalez, McKay and Mvondo, 2001). Some investigators propose cyanobacteria were equipped early on (at least 2.7 Gyr ago) with the nitrogenase system (Nisbet and Sleep, 2001; Berman-Frank, Lundgren and Falkowski, 2003). However, it remains to be determined exactly when cyanobacteria acquired nitrogenase, and whether it was vertically inherited from the last common ancestor of life's three domains, acquired by horizontal gene transfer from methanogenic Archaea, or evolved through some other process (reviewed by Raymond et al., 2004).

If it is true that "fixed nitrogen, not phosphorus, limits primary productivity on geological timescales" (Falkowski, 1997), then nitrogen would have been the nutrient most critical for the widespread proliferation of cyanobacteria required to produce abundant amounts of oxygen. But even with an ability to fix nitrogen, cyanobacteria could still have been limited by a scarcity in abiotic sources of fixed nitrogen. Indeed, the commonly held assumption that biological nitrogen fixation necessarily precludes nitrogen limits on the expansion of life is questionable (Anbar and Knoll, 2002).

In this context it is pertinent that nitrogen fixation is one of the most energetically expensive processes in biology (Raymond et al., 2004). Nitrogenase requires 16-24 ATP molecules per $N_2$ fixed to provide the dissociation energy of 940 kJ mol$^{-1}$ needed to break the extremely stable $N_2$ triple bond and subsequently produce ammonia (Lide, 1997; Fujita and Bauer, 2000; Catling et al. 2005). Given the energy demands of the nitrogenase system, there was probably a selective advantage to "packaging" it within cyanobacterial cells or cell aggregates that also generate large amounts of ATP through the photosynthetic light reactions. While this arrangement could provide cyanobacteria with a more reliable source of fixed nitrogen, it would not be cheap. It probably would have allowed these cells to gradually proliferate during the late Archean and early Proterozoic, but hardly enabled "blooms." This is reflected by the apparent lack of evidence for blooms in the geochemical record of this period (Anbar and Knoll, 2002). Even most of the modern open ocean remains impoverished of biologically usable nutrients, including nitrogen, despite the presence in these waters of diazotrophs. This constrains microorganism growth and requires survival strategies adapted to an oligotrophic environment (Zehr, Carpenter, & Villareal, 2000).

Marine cyanobacteria of the late Archaean and early Proterozoic most likely encountered similar or even greater nutrient constraints, and probably most closely resembled the "survivalist" growth model proposed for certain modern phytoplankton. Marine microorganisms which utilize this strategy have extensive "resource-acquisition machinery," consisting of abundant photosynthetic pigments and proteins, as well as other enzymes, and this results in these cells having a very high nitrogen-to-phosphorus ratio ($> 30$). This strategy enables slow growth in resource-poor environments where microorganisms employing other strategies fail to compete (Arrigo, 2005).

Despite the energetic advantages, packaging nitrogenase within oxygen-producing cyanobacteria creates complications because nitrogenase is inhibited by oxygen. As a result, cyanobacteria have had to evolve various mechanisms, often rather complex, to solve this problem (Capone et al.,1997; Berman-Frank, Lundgren and Falkowski, 2003; Raymond et al., 2004). It is possible the need to evolve and maintain these non-trivial adaptations to protect nitrogenase from oxygen could have also had the effect of slowing the proliferation of cyanobacteria throughout the oceans (Anbar & Knoll, 2002).

For example, some marine cyanobacteria such as unicellular *Cyanothece* fix nitrogen only at night (Berman-Frank, Lundgren and Falkowski, 2003). While this allows the nitrogenase in these cells to avoid the oxygen generated during the day by photosystem II, it also temporally restricts nitrogen fixation and hence limits the amount of fixed nitrogen produced. A second type of adaptation which has evolved in some filamentous cyanobacteria, such as the pelagic species *Anabaena gerdii*, is cellular differentiation. In this case nitrogenase expression is confined to specialized cells called heterocysts. These cells form a micro-anaerobic environment by virtue of a thickened cell wall and membrane which slow the inward diffusion of external oxygen, and they also do not express oxygen-generating photosystem II (Berman-Frank, Lundgren and Falkowski, 2003; Madigan, Martinko and Parker, 2003). Because heterocysts are a small minority among the cells comprising filamentous cyanobacteria, this spatial restriction to nitrogenase activity, like the temporal restriction described above, limits the amount of fixed nitrogen produced.

Yet another factor to consider in evaluating the effectiveness of nitrogenase in surmounting nitrogen limitations is the different specific activities of various forms of this enzyme. Nitrogenase is a metalloenzyme, and the three known forms require either molybdenum, vanadium, or iron at a particularly important catalytic site. At present the most common form of the enzyme, which is present in all known modern diazotrophs, uses molybdenum. Its specific activity is about 1.5 times that of the form which uses vanadium (at $30^{o}C$), and it is at least this much more active than the form which uses iron. As molybdenum was probably scarce in Archean and Proterozoic oceans, whereas iron was relatively abundant, the less efficient iron form of nitrogenase probably accounted for most of the fixed nitrogen supply (Anbar and Knoll, 2002; Berman-Frank, Lundgren & Falkowski, 2003).

Thus, several factors may have constrained the pace of cyanobacterial expansion and contributed to the lag between the emergence of cyanobacteria at least 2.7 Gyr ago and the first accumulation of atmospheric oxygen about 2.3 Gyr ago. These factors include: (1) the continued scarcity of fixed nitrogen from abiotic sources, (2) the high energy demands of the nitrogenase enzyme, (3) the need to evolve and maintain costly and complex adaptations to protect nitrogenase from oxygen, (4) the predominance of the less-efficient iron form of nitrogenase in Archean and Proterozoic oceans, and (5) the greater need of cyanobacteria for fixed nitrogen (discussed below). As a result of these factors, many millions of years could have transpired before the cyanobacterial biomass became the "large biological source" required to produce abundant oxygen. While the ability to fix nitrogen clearly facilitated the eventual widespread expansion of cyanobacteria, this expansion did not necessarily occur "instantly on a geological timescale," as some have proposed (Nisbet and Sleep, 2001).

Moving forward in time, another long period of nitrogen stress is thought to have prevailed during much of the mid-Proterozoic (at least 1.8 Gyr to 1.25 Gyr ago), and the factors creating this stress would have affected cyanobacterial proliferation despite their ability to fix nitrogen. This is because the redox-sensitive trace metals crucial to the catalytic activity of nitrogenase (such as molybdenum and iron) were probably scarce during this period as a result of the oceans being moderately oxic at shallow depths and euxinic (anoxic and sulfidic) at lower depths. These metal scarcities would have greatly curtailed nitrogen fixation by cyanobacteria and as a result global rates of $N_2$ fixation probably declined (Anbar and Knoll, 2002).

Severe limits on nitrogen fixation during the mid-Proterozoic, and by extension steady-state or perhaps even reduced primary production by cyanobacteria, may help explain the long period of stasis during the same period (roughly 2.0 Gy to 1.0 Gy ago) when apparently the increase in biospheric oxygen was relatively small or even nonexistent (Arnold et al., 2004; Canfield, 2005; Catling et al., 2005). It is likely the cyanobacterial population during this period was itself static (or even declined?), and, if so, the amount of oxygen it produced would have been similarly affected. This long period of stasis resulted in a significant increase in Earth's "oxygenation time" (Catling et al., 2005), and perhaps also greatly slowed the pace of eukaryotic evolution (Anbar and Knoll, 2002).

In the final analysis, it appears the ability to fix nitrogen by cyanobacteria was not a panacea for the problem of nitrogen limitations. But without it these organisms may have forever remained a minor life form, and Earth's oxygenation almost certainly would not have occurred.

### 3. Cyanobacteria and the Need for Fixed Nitrogen

The importance of fixed nitrogen for the multiplication of cyanobacteria may be greater than it is for most if not all other prokaryotic cells. Like other prokaryotes, cyanobacteria need fixed nitrogen for the biosynthesis of DNA, RNA, amino acids, various cell wall components, and compounds such as ATP and NADPH, which are critical to metabolic pathways (Madigan, Martinko and Parker, 2003). However, the unique composition of the photosynthetic apparatus in cyanobacteria would appear to create a substantially greater need for fixed nitrogen in several ways.

First, it is needed for the biosynthesis of the extensive protein components of photosystems I and II of the light reactions. Compared to all other photosynthetic prokaryotes, only cyanobacteria have both systems (Kuhlbrandt, 2001; Nisbet and Sleep, 2001).

Secondly, fixed nitrogen is essential for biosynthesis of the abundant photosynthetic pigments (chlorophyll *a* and phycobilins) and the proteins which hold them in place within the thylakoid membrane system (Kuhlbrandt, 2001; Madigan, Martinko and Parker, 2003) that dominates the interior of cyanobacteria (Fig. 1). Not only does each chlorophyll *a* molecule require four nitrogen atoms, but these atoms are a critical part of the porphyrin ring structure of the molecule, a source of excited electrons that produce ATP and NADPH in the light reactions. Chlorophyll *a* is required in abundance for both photosystems I and II (Jordan et al., 2001; Kulhbrandt, 2001; Zouni et al., 2001).

While it is the case that other photosynthetic prokaryotes synthesize some form of bacteriochlorophyll, cyanobacteria also need fixed nitrogen for the biosynthesis of phycobilin accessory pigments such as phycoerythrin and phycocyanin. These compounds are open chain tetrapyrroles which also require four nitrogen atoms per molecule. Among prokaryotes phycobilins are found only in cyanobacteria, and they play a major role in the harvesting of light energy. Other photosynthetic prokaryotes utilize carotenoid accessory pigments, but these compounds do not require nitrogen (Madigan, Martinko and Parker, 2003).

Thirdly, cyanobacteria are also unique in possessing a structure called cyanophycin, which consists of a polymer with equimolar amounts of L-aspartic acid and L-asparagine. Cyanophycin is a nitrogen storage product reported in almost all species and it may comprise as much as 16% of the cell dry weight. It appears to be important for maintaining the structural integrity of the thylakoid membrane system, including pigment and protein components (Stephan, Ruppel and Pistorius, 2000). The presence of cyanophycin in cyanobacteria strongly suggests that these organisms have a greater need for nitrogen than other prokaryotes. If so, it would underscore the importance of their ability to fix nitrogen in the absence of sufficient abiotic sources of this nutrient. However, a greater need for fixed nitrogen would have imposed constraints on the rate of cyanobacterial proliferation – constraints that other prokaryotes would not have experienced.

### 4. A Genetic Link Between Nitrogen Fixation and Photosynthesis

Another important way in which the evolution of nitrogen fixation may be linked to the oxygenation of Earth's biosphere, and perhaps to the evolution of photosynthesis as well, is revealed by the structural and functional similarities between subunits of the nitrogenase enzyme system and subunits of certain enzyme complexes involved in chlorophyll and bacteriochlorophyll biosynthesis.

Two enzymes important in (bacterio)chlorophyll biosynthesis, which have been studied in this respect, are protochlorophyllide reductase and chlorophyllide reductase. The first catalyzes an essential late step in the biosynthesis of both chlorophyll and bacteriochlorophyll, while the second catalyzes an essential late step in the biosynthesis of bacteriochlorophyll (Burke, Hearst, and Sidow, 1993; Raymond et al., 2004). Both enzymes are comprised of three subunits, and in each case all three subunits exhibit notable amino acid sequence similarity to certain subunits that comprise the nitrogenase enzyme system.

For example, in the purple nonsulfur bacterium *Rhodobacter capsulatus* (which conducts anoxygenic photosynthesis using a form of bacteriochlorophyll), the three genes encoding the protochlorophyllide reductase subunits are designated *bchL, bchN,* and *bchB*. The cyanobacterium *Plectonema boryanum* and the green alga *Chlamydomonas reinhardii* (which both conduct oxygenic photosynthesis using chlorophyll *a*) have very similar genes encoding protochlorophyllide reductase which are labeled *chlL, chlN,* and *chlB*. The amino acid sequences deduced from these genes show a significant similarity between the BchL/ChlL, BchN/ChlN, and BchB/ChlB subunits of protochlorophyllide reductase and the sequences of the NifH, NifD, and NifK subunits of nitrogenase, respectively, which have been determined from a variety of diazotrophs. The sequence similarity between protochlorophyllide reductase and nitrogenase is greatest between the BchL/ChlL and NifH subunits. In this case there is a 33% overall identity and a 50% similarity (Fujita and Bauer, 2000). Furthermore, the sequence similarity between the subunits of protochlorophyllide reductase and nitrogenase also translates into very similar molecular architectures and catalytic functions for each enzyme complex (Burke, Hearst, and Sidow, 1993; Fujita and Bauer, 2000).

Nitrogen fixation and photosynthesis are both thought to be ancient innovations (Xiong et al., 2000; Raymond et al., 2004). Which evolved first is not yet firmly established, but the phylogenetic distribution of nitrogen fixation among both Archaea and Bacteria strongly suggests that it is the more ancient process and therefore preceded photosynthesis (Berman-Frank, Lundgren, and Falkowski, 2003; Raymond et al. 2004). In contrast, true photosynthesis is found only in Bacteria (Woese, 1987). Furthermore, amino acid sequence analyses of protein subunits comprising nitrogenase in diverse Archaea and Bacteria have led some investigators to conclude that multiple nitrogenase genes were already present in the last common ancestor (LCA) of life's three domains (Fani, Gallo, and Lio, 2000). This also strongly suggests that nitrogen fixation arose before photosynthesis.

Moreover, the significant structural and functional similarities between the subunits of nitrogenase and the reductase enzymes involved in (bacterio)chlorophyll biosynthesis have led others to propose that an ancient duplication of a gene encoding the iron protein subunit of a primitive nitrogenase gave rise to the first form of protochlorophyllide reductase (Burke, Hearst, and Sidow, 1993). Additional gene duplication and subsequent evolution led to the recruitment of protochlorophyllide reductase and chlorophyllide redectase into the pathways for (bacterio)chlorophyll biosynthesis (reviewed by Raymond *et al.*, 2004). Thus, at this point, the phylogenetic and molecular evidence indicate nitrogen fixation arose before photosynthesis, but more work on this question is needed.

Some form of (bacterio)chlorophyll is obligatory for true photosynthesis to take place (Madigan, Martinko and Parker, 2003). If it is the case that genes encoding enzymes critical for (bacterio)chlorophyll biosynthesis were originally derived from a gene encoding a primitive nitrogenase, then this indicates photosynthesis may not have evolved if nitrogen fixation had not evolved first. Thus, oxygenation of Earth's biosphere by cyanobacteria would then appear to be contingent on an earlier "nitrogen revolution" that not only provided the ability to fix nitrogen necessary for the proliferation of these organisms, but also some of the genetic raw material that made it possible for (bacterio)chlorophyll-based photosynthesis to arise. To the extent that the physical and chemical properties of (bacterio)chlorophyll (such as its light absorption properties, geometric rigidity, and great stability) make it a "universal pigment" (Raven & Wolstencroft, 2004), we might predict the scenario described here connecting nitrogen fixation to photosynthesis and biospheric oxygenation may also play out on other Earth-like planets.

## 5. Conclusions and Future Directions

It is concluded that Precambrian cyanobacteria would not have been able to oxygenate Earth's biosphere if their emergence had not been preceded by the evolution of nitrogen fixation, and if these organisms had not also acquired the ability to fix nitrogen at the beginning of or very early in their history. Furthermore, the evolution of nitrogen fixation may also have been a precondition for the evolution of (bacterio)chlorophyll-based photosynthesis. These hypotheses will be tested as additional genomic, biochemical, and geological evidence sheds more light on the evolution of nitrogen fixation, photosynthesis, and cyanobacteria. Accordingly, the following lines of enquiry are proposed as ways of testing these hypotheses.

(1) Various types of fossils have been used to date the emergence of cyanobacteria by 2.7 Gyr ago and perhaps earlier. These include molecular (hydrocarbon biomarkers) fossils, microbial biofilm and stromatolite fossils, and filamentous microfossils (Nisbet and Sleep, 2001; Canfield, 2005; Catling et al., 2005). Further studies of the distribution of these fossil types in the geologic record will help elucidate the expansion of cyanobacteria through time, and thereby provide some idea of the rate at which they proliferated and enlarged their range.

Additional insight into the cyanobacterial component of the Bacterial population at various points in the geologic record may also be obtained by measuring the 2-methylhopane (2-MHP) index, which is a ratio of the abundance of cyanobacterial biomarkers to more general Bacterial biomarkers (Xie et al., 2005). Surveys of the 2-MHP index in a large sample of late Archean and Proterozoic sediments should tell us much about the early history of Earth's microbial community structure, and further refine our understanding of the proliferation and expansion of cyanobacteria after they first appeared.

(2) As mentioned earlier, the catalytic activity of nitrogenase requires certain metal cofactors. All forms of nitrogenase require at least 34 iron atoms for each fully assembled nitrogenase molecule. Depending on the particular form of nitrogenase, each fully assembled molecule also requires either two molybdenum atoms, two vanadium atoms, or two additional iron atoms (Fujita and Bauer, 2000; Raymond et al., 2004).

Nitrogenase makes up about 10% of the total cellular proteins in many diazotrophs (Berman-Frank, Lundgren, and Falkowski, 2003). Thus, it may be possible to infer the existence of nitrogenase in ancient cyanobacteria by correlating the presence of various fossil types described in (1) above with an enrichment within the same fossil samples of iron, molybdenum, and/or vanadium. In this context, it is noteworthy that other investigators have recently proposed studying redox-sensitive metal abundances in sediments as a way of obtaining more paleobiological data from the geologic record (Anbar and Knoll, 2002). Of course, this approach will be complicated by the existence of other cellular sources of these metals, such as the iron-containing proteins that are part of the photosystems I and II (Sunda and Huntsman, 1997). In addition, enrichment for these metals can only be claimed if comparisons to carefully selected control Bacterial fossils of known non-diazotrophic species are used as a baseline. A parallel approach that measures the comparative iron, molybdenum, and vanadium content of extant diazotrophic and non-diazotrophic cyanobacterial species might also help establish baselines for determining metal enrichment due to the presence of nitrogenase.

Finally, perturbations of the nitrogen cycle influenced by the widespread occurrence of nitrogenase in ancient marine ecosystems might be dated by an analysis of nitrogen isotopes (Holland, 2002) and/or the development of a reliable proxy for these isotopes (Anbar and Knoll, 2002). Because cyanobacteria are the organisms mainly responsible for fixing nitrogen in marine ecosystems (Kasting and Siefert, 2002), correlating nitrogen cycle perturbations over geologic time with fossil evidence for cyanobacteria may be another way of establishing when these organisms gained the ability to fix nitrogen.

(3) Testing whether or not cyanobacteria have a greater need for nitrogen compared to other prokaryotes should be straight forward with respect to the required chemical methodology. It is well established that a typical bacterial cell is about 12% nitrogen by dry weight (Madigan, Martinko, and Parker, 2003). To adequately test the hypothesis, this parameter needs to be measured in a variety of cyanobacterial species as well as an adequate sample of heterotrophic and other autotrophic species which would serve as controls for making meaningful comparisons.

It would be especially important for the autotrophic controls to include other photosynthetic prokaryotes such as green sulfur bacteria, green nonsulfur bacteria, purple bacteria, and heliobacteria. These photosynthetic organisms have bacteriochlorophyll-based pigment systems, but lack the nitrogen-containing phycobilin pigments of cyanobacteria. They also contain either photosystem I or II, but not both, as is the case in cyanobacteria. Interestingly, many of these other photosynthetic prokaryotes, which are considered more ancient than cyanobacteria, also have the ability to fix nitrogen via the nitrogenase system (Des Marais, 2000; Xiong et al., 2000; Kuhlbrandt, 2001; Madigan, Martinko, and Parker, 2003).

(4) As discussed above, the available evidence indicates nitrogenase existed in organisms that predate cyanobacteria. Accordingly, two possible scenarios postulate nitrogenase genes (usually tightly linked within highly conserved operons) encoding the core protein subunits of the holoenzyme were either inherited by cyanobacteria from the last common ancestor of life's three domains, or were acquired by horizontal transferred from methanogenic Archaea (reviewed by Raymond et al., 2004). The latter are thought to be more ancient than cyanobacteria (Woese, 1987).

However, the molecular data used to construct these scenarios are somewhat problematic because of confounding factors such as gene duplication, loss, recruitment, fusion, sequence divergence, and horizontal gene transfer (Raymond et al., 2004). Similar to (2) above, the study of certain redox-sensitive metal abundances in fossil Archaea and other similarly ancient fossils might provide more direct evidence for the presence of nitrogenase in these very early life forms, and may even permit the establishment of a firm date for the appearance of nitrogenase. As is the case for cyanobacteria, unique molecular fossils (biomarkers) have also been identified for Archaea (Kuypers et al., 2001). In the absence of Archaeal microfossils, perhaps the occurrence of these molecular fossils along with enriched levels of metals such as iron, molybdenum, and vanadium may help establish a date for the presence of nitrogenase in ancient Archaea. If these and other methods are able to demonstrate that nitrogenase predated that appearance of photosynthetic Bacteria, it would lend further support to the hypothesis that the evolution of nitrogen fixation was a precondition for the evolution of (bacterio)chlorophyll-based photosynthesis.

## Acknowledgements

The author is grateful to Rocco Mancinelli, Joan Depew, Allan Sandage and Kanniah Rajasekaran for critically reading the manuscript and providing helpful comments. Klaus-Peter Michel and Elfriede K. Pistorius kindly provided the *Synechocystis* photo (Fig. 1).

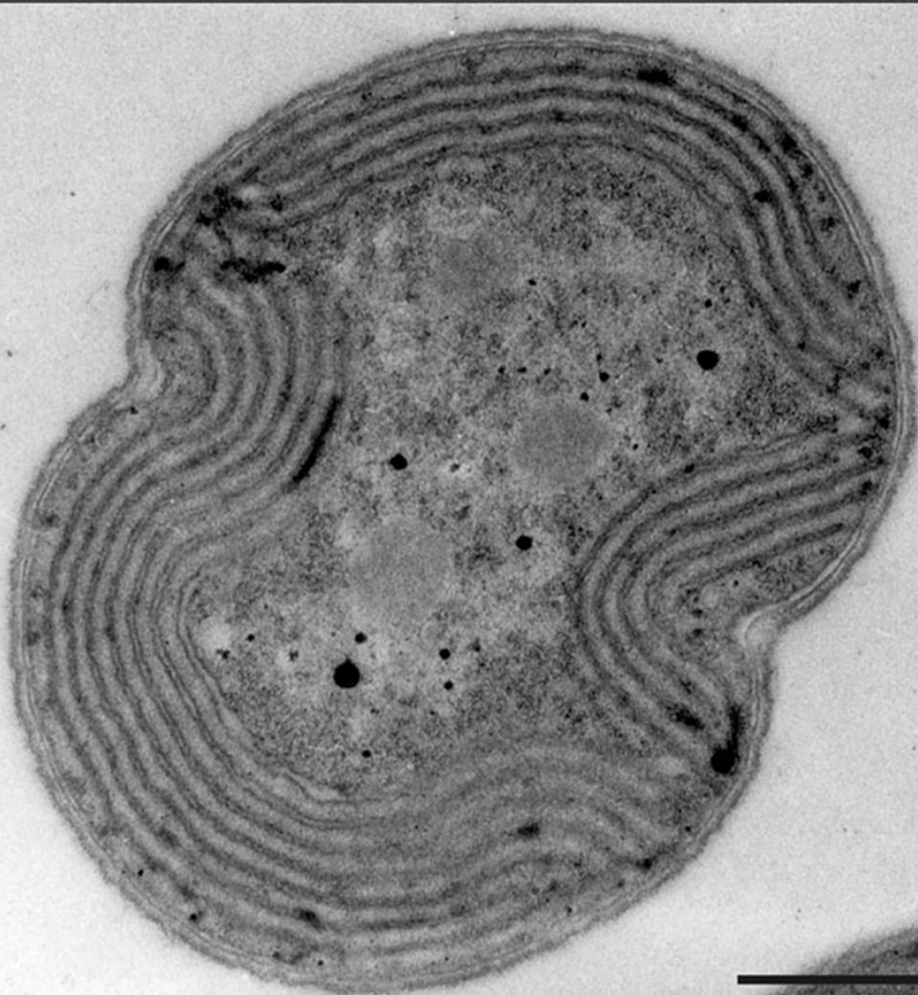

Fig. 1. A unicellular cyanobacterium of the *Synechocystis* group, which includes marine diazotrophs (Zehr *et al.*, 2001). *Synechocystis* and other unicellular cyanobacteria make a major contribution to nitrogen fixation in modern oligotrophic oceans (Montoya et al., 2004). Light harvesting antennae and light reaction centers replete with photosynthetic pigments and proteins are embedded within thylakoid membranes that ring this cell in 4-7 layers. Scale bar, 1.0 µm. Photo credit: Drs. D. P. Stephan and E. K. Pistorius, Univ. Bielefeld, Germany (Stephan, Rupel and Pistorius, 2000).